# Transforming Computational Lithography with AC and AI – Faster, More Accurate, and Energy-efficient

Saumyadip Mukhopadhyay[1]*, Kiho Yang[2], Kasyap Thottasserymana Vasudevan[1], Mounica Jyothi Divvela[1], Selim Dogru[1], Dilip Krishnamurthy[1], Fergo Treska[2], Werner Gillijns[2], Ryan Ryoung han Kim[2], Kumara Sastry[1], Vivek Singh[1]

[1] NVIDIA              [2] IMEC

From climate science to drug discovery, scientific computing demands have surged dramatically in recent years – driven by larger datasets, more sophisticated models, and higher simulation fidelity. This growth rate far outpaces transistor scaling, leading to rising costs, energy consumption, and emissions that are increasingly unsustainable. Semiconductor manufacturing is no exception. Computational lithography – involving transferring circuitry to silicon in diffraction-limited conditions – is by far the largest workload in semiconductor manufacturing. It has also grown exceptionally complex as miniaturization has advanced in the angstrom-era, requiring more accurate modeling, more intricate corrections, and broader solution-space exploration.

Accelerated computing (AC) offers a solution by dramatically freeing up the compute and power envelope. AI augments these gains by serving as high-fidelity surrogates for compute-intensive steps. Together, they present a sustainable, next-generation computing platform for scientific workloads. This new paradigm relies on a heterogeneous architecture, where control- and data-intensive steps are routed to the most efficient devices – warranting a fundamental redesign of the software stack. For computational lithography, NVIDIA cuLitho reinvents the core primitives – diffractive optics, computational geometry, multi-variant optimization, data processing – to achieve a transformative greater than 50X end-to-end acceleration.

Beyond dramatically faster cycles, this expanded compute envelope enables more rigorous solutions, including curvilinear masks, high-numerical aperture extreme ultraviolet (high-NA EUV) lithography, and subatomic modeling. We reinvest a small fraction of the freed-up compute to include through-focus correction for better process resilience. Silicon experiments at IMEC show significant benefits compared to conventional methods – 35% better process window and 19% better edge placement error. This is the first quantified chip-scale demonstration of the lithography benefits of AC and AI in silicon.

## 1. INTRODUCTION

Computational lithography has been the cornerstone of chipmaking. Ever since its inception in the nineties, it has continued to bridge the gap between intended design and reality. It has evolved from simple rules to physics-guided decorations [1,2], and to more recently physics and optimization driven intricate patterns in Inverse Lithography Technology [3,4]. This complexity is fueled by the relentless advance of process nodes, as captured by Moore's law – the industry metronome that predicts transistor count on a chip to double every two years. The resulting computational burden of lithography has grown exponentially, primarily driven by growing mask complexity, stricter process requirements, and increasing need for more precise modeling.

*1. Increasing mask complexity:* Modern chips comprise tens of billions of transistors and trillions of layout polygons, all of which undergo physics-based optical and resist simulation, model-based correction, and inversion to achieve accurate wafer patterns [5]. Early generations of optical lithography required little or no proximity correction because features were large relative to the imaging wavelength. However, beginning around the 22-nm node, the interaction between closely spaced features caused severe optical proximity effects. To maintain pattern fidelity, foundries increasingly relied on multiple patterning techniques – double, triple, and even quadruple patterning [6,7]. This shift dramatically increased the complexity and number of masks required per design layer. At advanced nodes (7 nm and below), masks have migrated from Manhattan geometries toward curvilinear mask shapes, which provide superior imaging performance by reducing diffraction artifacts [8]. Inverse Lithography

Technology (ILT) represents the next phase of this evolution - producing fully continuous, curvilinear mask geometries derived from rigorous optimization that maximizes process window, depth of focus, and overall edge-placement accuracy [9,10].

*2. Stricter process requirements:* As feature sizes approach or fall below the illumination wavelength, depth of focus collapses, and both line-edge roughness (LER) and stochastic variability become critical [11,12]. Meeting accuracy requirements requires finer simulation grids, more elaborate resist and optical models (e.g., vector imaging, resist blur kernels, secondary electron effects), and tighter convergence criteria (Mack, 2006) [2]. These refinements dramatically expand computational demand - OPC and ILT workloads have grown superlinearly with each process generation [13].

*3. More accurate modeling:* Traditional scalar imaging and thin-mask approximations have broken down since early 2000s, due to effects induced by mask topography [14-16]. With sub-wavelength mask features and depth of etched openings acting as scattering objects, mask topography induces polarization and phase dependent effects on the diffracted light [17-20]. These effects are more pronounced in off-axis illumination and high-NA DUV systems [21-23]. Failure to capture these effects results in critical patterning defects. With rigorous mask modeling computationally prohibitive (thus far), several compact mask models – from library-based domain decomposition techniques [24-27] to boundary layer models [28] – have been developed to capture mask-topography effects. At sub-2nm nodes, assumptions behind these compact models are slowly breaking down, generating interest in chip-scale rigorous diffractive solvers [29,30].

*Electronic mail: saumyadipm@nvidia.com

*4. EUV lithography*: EUV introduces stronger 3D mask effects, including shadowing, absorber edge rounding, and polarization-dependent phase errors [31]. Additionally, the low photon counts (inherent to EUV) introduce stochastic photon shot noise, making accurate simulation both computationally intensive and inherently probabilistic [11,32]. These challenges intensify as the industry moves toward high-NA EUV scanners, where reduced depth of focus, increased aberration sensitivity, greater anamorphic magnification, and steeper ray angles amplify 3D effects and demand even more rigorous simulation fidelity [33].

Computational lithography is already the biggest workload in semiconductor manufacturing, and the continued process innovation is rapidly pushing the compute envelope [34]. Transistor scaling hasn't been keeping pace with these growing needs. Energy use has grown even faster since the breakdown of Dennard scaling around 2005, which states smaller transistors are faster and more energy-efficient [35]. This enabled each new generation of processors to use twice the number of transistors – and realize the resulting performance improvements – without an increase in power consumption. As devices shrank further, quantum tunneling and leakage currents nullified the gains from power scaling, causing energy use to increase with each generation [36,37]. This has made power efficiency a central concern in computing.

For energy-aware computing, heterogeneous systems present a way forward. They include serial and parallel processing devices, with tasks in an application mapped to the best processing device available in the system. The presence of multiple devices allows programs to utilize concurrency and parallelism, improving both performance and power. In recent years, heterogeneous computing has become much more important due to stagnant single-threaded performance. From about 1986 to 2003, the single-thread performance of CPUs increased by more than 50% per year on average, but since 2003, the improvement in single-thread performance has decreased, to the point that from 2015 to 2017, it has been growing at less than 4% per year [38]. Over the same period, gains from multi-threaded and heterogeneous systems (such as GPU paired with CPU) have far outpaced single-threaded scaling.

Early efforts to accelerate computational lithography used field-programmable gate arrays (FPGAs) for imaging acceleration, exploiting the regular structure of FFT-based convolutions [39]. Graphics processing units (GPUs) soon proved far more effective, offering far higher throughput, memory bandwidth, and programmability [40,41]. Several review papers summarize prior GPU-based work in OPC [42-44]. To date, most GPU acceleration in computational lithography has focused on GPU-friendly components, such as OPC simulations [45] or pixel-based ILT algorithms [40,46]. Because only parts of the flow are accelerated, end-to-end speedup is limited. For example, Zheng et al. (2024) [47] reported a 5.3X speedup for a traditional edge-based OPC flow by offloading simulation alone to GPUs. While significant, this gain does not offset the projected compute growth at advanced nodes. Their flow also excludes mask rule checking (MRC) and Boolean geometry operations, which are essential in production OPC. Incorporating these modules without GPU acceleration would further erode end-to-end speedup.

Pixel-based ILT has benefited from significant GPU acceleration, most notably in the work of Pang et al. [42,48]. ILT generates lithographically superior, curvilinear masks through rigorous optimization, but its computational cost is typically an order of magnitude higher than that of traditional OPC [42,44]. Using GPUs, Pang et al. [48] reduced ILT runtimes from weeks to days for DUV lithography. Even so, runtimes remain too long for widespread adoption in high-volume manufacturing that use EUV lithography [44,49,50]. The significantly larger ILT runtime for EUV is partly due to the finer modeling grids required (Pang, 2021) [42]. Increased geometric complexity arising from the curvilinear nature of ILT masks is another contributor to longer runtimes [51]. Indeed, Peng (2022) [52] identified geometry processing as a major ILT bottleneck, highlighting the need for GPU acceleration. Although Pang et al. (2019) [48] included MRC enforcement in their ILT engine, it is unclear whether geometric operations were GPU accelerated. As curvilinear mask representations evolve toward more complex formats, such as multigons composed of parametric curves [51,53], the computational cost of geometry processing is expected to increase further.

Taken together, these observations indicate that end-to-end GPU acceleration of both OPC and ILT is essential for sustaining patterning capability at advanced technology nodes. The fundamental challenge in achieving end-to-end acceleration for any software system, including computational lithography, is the limitation imposed by Amdahl's law. Nearly sixty years ago, Gene Amdahl observed that speedup is limited by the portions of a program that is not accelerated (Amdahl, 1967) [54]. Once one part is sped up, another becomes a new bottleneck, capping overall gains. Further acceleration requires tackling the new bottleneck – and then the next. In OPC, GPU-friendly imaging is typically the dominant bottleneck often accounting for half the runtime. However, even with infinite acceleration of imaging, Amdahl's law limits the overall OPC speedup to 2X (Singh, 2023) [34]. Unlocking additional speedup hinges on accelerating the remainder – primarily geometric tasks like polygon Booleans, spatial queries, mask rule enforcement, contour extraction, measurements, and OASIS processing.

However, accelerating computational geometry on GPUs is challenging. GPUs deliver peak performance when workloads exhibit high data parallelism, coalesced memory access, and minimal control-flow divergence – characteristics well-suited to their SIMT architecture (NVIDIA CUDA C Programming Guide; CUDA Best Practices Guide). In contrast, geometric operations often involve uncoalesced access and control-flow divergence, making them harder to accelerate efficiently on GPUs [55-57]. For example, edge intersection checks, a fundamental step in polygon Booleans often result in uncoalesced memory access, as intersecting edges are rarely stored within the same cache line.

Through advanced GPU algorithms, cuLitho addresses the long-standing challenge of end-to-end acceleration in computational lithography, achieving speedups as large as 57X over production CPU flows. The remainder of this paper is organized as follows. We first describe cuLitho's acceleration of major OPC building blocks and review prior efforts on parallelizing computational geometry algorithms. We then provide a detailed overview of AI methods in OPC and explain how cuLitho incorporates these techniques to deliver additional speedup. Next, we present our approaches for through-focus correction and post-OPC silicon validation. This is followed by detailed results on OPC acceleration, pre-silicon verification, and silicon metrology. Finally, we discuss these results in the broader context of sustainable computing and conclude the paper.

## II. ACCELERATING OPC BUILDING BLOCKS USING CULITHO

### A. Booleans

Boolean operations such as AND, OR, NOT, HEAL, XOR, SUB, SIZE, and TOUCH are critical to the layer preparation step in OPC recipes. Advanced OPC requires a scalable and accurate Boolean engine capable of handling arbitrary input geometry - including non-Manhattan, non-convex, hole-containing, self-intersecting, and degenerate polygons [57,58]. All Boolean operations involve 3 fundamental steps: 1) calculation of edge intersection vertices, 2) determining which vertices needs to be included in the output, and 3) constructing the output polygons by linking vertices in the correct order [59,60].

Early boolean algorithms were focused on specific cases of polygon clipping – a type of AND – on CPUs [60]. For example, Sutherland and Hodgeman's (1974) [61] algorithm is limited to convex clip polygons and Liang and Barsky (1983) [62] supports only rectangular clip polygons. Other methods [64-68] do not support sell-intersecting polygons. A more general algorithm by Greiner and Hormann (1998) [60] can handle all but degenerate polygons [59].

Only a few algorithms in the literature can handle arbitrary polygons with degeneracies [61]. They include Weiler's (1980) [70] graph-based method, and more efficient, plane-sweep based methods by Vatti (1992) [59] and Martinez et al. (2009) [71]. Plane-sweep reduces the asymptotic complexity of edge intersection finding from $O(n^2)$ to $O((n+k) \log(n))$ where n and k are edge and intersection counts respectively [71]. The core idea is to transform a 2D geometric problem into a sequence of simple quasi-1D updates. After sorting all edge endpoints vertically, the algorithm scans the layout by moving a horizontal beam or a line upward. At each moment, a dynamically updated list of "active" edges that interact with the scan beam or line is maintained. The active edge list is kept ordered horizontally such that intersection needs to be checked only between adjacent edges. This allows the algorithm to avoid costly all-pairs intersection checks. As the sweep progresses, the method also determines which vertices contribute to the output and assembles them into polygons [59,71].

The most compute-intensive step in booleans is the edge intersection finding. Plane-sweep methods described above tackle this efficiently on CPUs. However, they do not translate directly to the SIMT architecture of GPUs due to their sequential logic and dynamic data structures [69,72,73]. In general, efforts to accelerate intersection computations on GPUs fall into three main categories: (1) parallel brute-force approaches, (2) parallel plane-sweep algorithms, and (3) methods based on spatial indexing structures.

The early work by McKenney et al. (2011) [72] focused on a naïve parallelization of brute-force, all-to-all, edge intersection checks on the GPU. In this method, each GPU thread independently computes the intersection between a pair of edges. While being "embarrassingly" parallel, the $O(n^2)$ complexity limits the scalability of this method [72,73]. Subsequent research has sought to partially mitigate this limitation by filtering out non-intersecting edge pairs using bounding-box overlap tests instead of rigorous intersection checks [69,74].

Although parallelizing plane-sweep algorithms on GPUs is inherently challenging, multiple studies have explored this direction. Paudel and Puri (2018) [55] proposed a hierarchical sweep strategy in which each sweep step initiates new sweeps from detected intersection points. This approach requires an iterative execution flow to capture all edge intersections. For densely intersecting layouts, however, the number of iterations can grow rapidly, and degrade performance. Frye and McKenney (2025) [75] improved upon this using CUDA Dynamic Parallelism, allowing parent threads to launch child kernels for intersection computation. However, since current GPUs can execute only up to 128 kernels concurrently (CUDA prog. guide), excessive child launches can lead to kernel scheduling overheads, increased execution latency, and resource contention.

Spatial-index based methods partition either the space or the data into smaller chunks for fast geometric searches [76]. The simplest spatial index is the uniform grid which partitions the space into equally sized cells. Layout edges are then mapped to cells of a uniform grid, and intersection checks occur only among edges within the same cell [73]. The cells, in turn, are processed in parallel across GPU threads, reducing the computational complexity from quadratic to near-linear. However, non-uniform edge densities can lead to workload imbalance, as dense regions take longer to process [55,75,76]. Adaptive grid refinement has been proposed [77], but the optimal grid size depends on the layout, and long edges spanning multiple cells lead to redundant work [75]. These issues make the approach less practical for OPC layouts with highly variable geometries [76].

To overcome these limitations, hierarchical spatial-indexing methods such as bounding volume hierarchies (BVHs) have been proposed [76]. In a BVH, geometric objects are recursively grouped into spatially coherent bounding volumes, with leaf nodes referencing individual objects and internal nodes enclosing their children. This hierarchy enables fast intersection queries with logarithmic complexity by pruning large portions of the search space [76]. As detailed in the next section, BVH is one of the most efficient spatial-indexes on the GPU [78].

Once intersection processing is accelerated, output polygon construction becomes the new bottleneck, especially for large data sizes [57]. Unlike serial plane-sweep algorithms [59], where polygon contours are progressively built during edge processing, parallel reconstruction cannot rely on sequential contour tracing. It often requires costly point-in-polygon queries – commonly implemented via ray-shooting – to classify vertices and assemble consistent output boundaries [57,73,75].

Booleans in OPC are more challenging than those in geospatial applications, which were the primary focus of many earlier efforts. In production OPC, Boolean algorithms must be accurate, robust, and efficient over trillions of complex, arbitrary polygons. Non-Manhattan layouts further increase both accuracy and performance demands. Finite-precision errors in intersection calculations can introduce spurious intersections, requiring refinement steps such as snap rounding to maintain geometric consistency (Hobby, 1999) [79]. Moreover, the use of double-precision arithmetic, required for non-Manhattan geometries, increases GPU register pressure. This can lead to reduced warp occupancy and overall performance [80,81].

cuLitho addresses the accuracy, scalability, and performance challenges of Booleans described above. Our boolean engine has been validated and benchmarked at scale against leading CPU implementations, achieving an average 34X performance gain.

In addition to the elementary operations of AND, OR, NOT, and XOR, cuLitho supports 48 other Boolean operations such as sizing and touch on both closed-chain polygons and open edge-chains. Furthermore, 32bit, 64bit, and double precision edge coordinates are also supported.

## B. Spatial queries

Many other OPC algorithms also rely on fast geometric search. For example, mask rule enforcement requires neighbor queries to identify pairs of edges that may violate minimum spacing or width rules [82,83]. Another example is layout rasterization, which involves containment checks to determine whether a pixel lies inside layout polygons or not

[84,85]. Amenable to fine-grained parallelism, BVHs have become ubiquitous in other GPU-accelerated applications such as ray-tracing [86], collision detection [87], and distance queries [88].

One of the most GPU-friendly search trees [89] is the linear-BVH (LBVH). The main idea behind LBVH is to reduce the construction process to a sorting problem along a 1D curve. For this set of geometric primitives in the 2D space are first mapped onto a 1D array using a space-filling Morton curve. This curve encodes the position vectors of primitive centroids into scalar Morton keys. Primitives are then sorted based on their Morton keys and the sorted set is then partitioned into a hierarchical structure. With fast GPU-based radix sort routines, good quality LBVH can be constructed on GPUs very efficiently [89].

cuLitho uses binary radix BVH which is an LBVH with at most 2 child nodes for each parent [92]. While more sophisticated LBVHs have been proposed to minimize the traversal time at the expense of higher construction costs [89], our BVH implementation provides a good overall balance between construction and traversal times in OPC layouts. The construction method is similar to Garanzha et al., (2011) [90]. We first compute bounding boxes and centroids of primitives in a CUDA kernel. Another kernel is used to compute the Morton keys using quantized centroids. This is followed by the sort step which uses the radix sort algorithm available in the NVIDIA CUB library. We then recursively partition the sorted primitive set top-down from the root-node to the leaves. At any given level, the nodes that need to be split are processed in parallel. The split-location is obtained using a binary search in the primitive sub-array at which the most significant Morton bit starts to differ. If all the Morton keys in the primitive sub-array collapse into one, then Morton codes in that node is requantized with a finer interval. Splitting is stopped when the primitive count in the node becomes equal to or less than the leaf-threshold. Finally, the bounding boxes of all nodes are populated upwards starting from the leaves.

For traversal, we use a stack-based, depth-first method, similar to the "while-while" algorithm proposed by Aila et. al., (2009) [91]. This method features two loops per query region (or ray), with different query regions processed in parallel by GPU threads. The first loop traverses internal nodes using a stack and until a leaf node is hit. The second loop performs intersection checks between the query region and primitives stored in the leaf nodes found by the first loop. While the depth-first strategy is one of the most efficient tree traversal methods on GPU, thread-divergence and irregular memory access patterns can cause latencies [78].

To maximize performance, we have implemented several optimizations. First, we maximize the warp occupancy by reducing the register pressure using shared memory and division-free calculations when possible. For range queries, we use a single-pass scheme to populate the search results, instead of a 2-pass scheme with first pass estimating the memory needed to store the results and the second pass populating the actual results [92]. With all the above optimizations, cuLitho spatial query is 165X faster than optimized CPU methods.

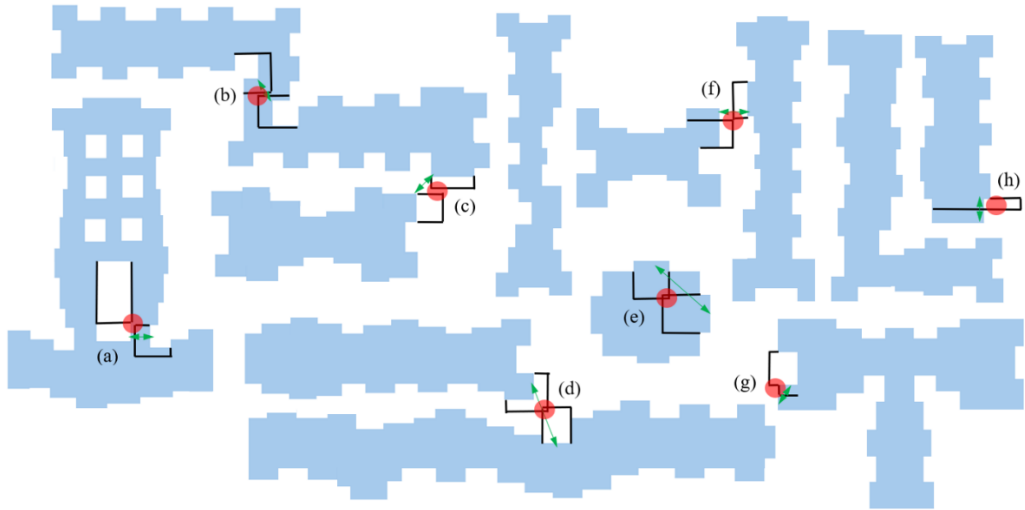

**FIG. 1**. Examples of dynamic layout configuration switches causing MRC violations. Inter-thread communication was needed in cuLitho MRC algorithms to avoid these. Shaded area indicates initial mask geometry, black lines show updated geometry after segment moves, green arrows indicate MRC clean initial state, and red circles show MRC violations after segment moves. (a) Int. E2E to Int. C2C, (b) Int. C2C to Int. E2E, (c) Ext. C2C to Ext. E2E (d) Ext. C2C from notch flip, (e) Int. C2C from nub flips, (f) Ext. E2E to Ext. C2C, (g) jog violation from notch flip, and (h) nub violation from jog flip

## C. Mask rule checks

Mask rule checks (MRC) ensure that corrected mask layouts remain manufacturable. Typical MRC rules constrain minimum edge-to-edge and corner-to-corner widths and spacings, notch and nub dimensions, jog lengths, and polygon areas [93]. In OPC, aggressive corrections often push mask geometries beyond these limits, producing MRC violations. These violations can either be repaired in a cleanup step after segment movement, or they can be prevented by constraining segment displacements during each OPC iteration. The latter approach is generally superior because post-repair cleanup may degrade pattern fidelity, especially in dense layouts [82,93,94].

In typical CPU implementations, segments are processed sequentially. For each segment, an MRC algorithm performs a neighbor search - often using a search tree - to identify nearby segments, then performs pairwise distance checks to determine whether proposed OPC moves would cause rule violations. The allowed displacement for each segment is the minimum permitted across all its neighbor interactions. Because edges are processed in order, CPU implementations exhibit traversal bias, which can produce asymmetric corrections even for symmetric patterns.

In culitho's GPU-accelerated MRC enforcement framework, a BVH-based neighbor query is executed in parallel for every OPC segment [95], followed by concurrent MRC checking and move limiting across all segment pairs. Parallelizing MRC introduces two key challenges absent in sequential CPU algorithms. First, when resolving an interaction with one neighbor, we must ensure that no new violations are created with other nearby edges being processed simultaneously in different GPU threads. Second, the algorithm must robustly handle geometric transitions between interaction types. For example, an external corner-to-corner interaction can become edge-to-edge if one participating edge moves forward far enough to pass the corner - potentially occurring in a different thread. In CPU implementations, such cases are straightforward to resolve because edges are processed sequentially, but on the GPU, they require careful coordinated sharing of intermediate state across threads. Examples of such cases are shown in FIG. 1.

The cuLitho MRC engine enforces all critical mask rules and delivers a 33X speedup over the CPU implementation on NVIDIA H200. Furthermore, the absence of traversal bias in the parallel algorithm preserves correction symmetry.

## D. Contour extraction

Lithographic contours are the level sets of the aerial image at the resist printing threshold. These contours are used for computing the OPC objective - the sum of all errors such as the edge placement error (EPE), pinch, bridge, pullbacks, and pushouts (De Bisschop, 2015) [96]. For extracting these contours from the aerial image, we employ the well-known "marching squares" algorithm, a 2D analog of the marching cubes method widely used in 3D visualization [97]. The marching squares algorithm works by systematically examining each cell in a discretized grid of the aerial image, determining where the printing threshold is crossed, and connecting these intersections to form continuous contour lines. One of the key advantages of this approach is that each cell can be processed independently, making it highly parallelizable and particularly suitable for GPU-based acceleration. After contours are generated, we perform BVH-based distance queries to compute OPC error metrics. Overall, our contour extraction and measurement pipeline achieves an average 176X speedup over CPU.

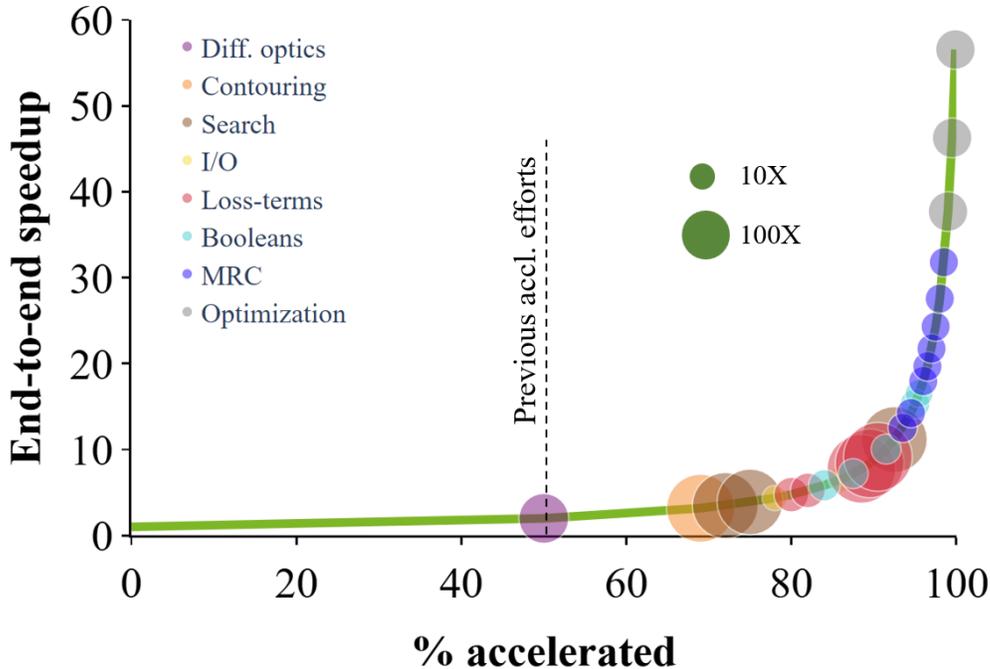

**FIG. 2.** Illustration of Amdahl's law for OPC. End-to-end speedup of a representative OPC flow versus the fraction of CPU execution time offloaded to the GPU. Circles represent individual GPU-accelerated OPC components, with the area proportional to the typical component-level GPU speedups achieved by cuLitho [100]. Previous acceleration efforts focused only on imaging stages accounting for approximately half of the total run time.

FIG. 2 shows the end-to-end speedup as a function of progressive GPU porting of OPC components by cuLitho. Achieving speedups on the order of 50X requires porting essentially all major modules to the GPU. Importantly, porting individual OPC modules alone does not guarantee end-to-end performance gains; efficient integration of GPU-resident components is critical. In particular, all the data needs to be kept in the GPU as long as possible to avoid host-device memory copy overheads. While GPU memory bandwidth is extremely high – 4.3 TB/s for NVIDIA H200 cards with HBM3e memory – the host–device interface is dramatically slower. PCIe Gen5 interconnect has only 128 GB/s bandwidth and the more advanced NVLink interconnect tops at 900 GB/s [98]. This disparity means that even a small number of host–device transfers can negate the performance gain from GPU acceleration. For unavoidable data transfers such as extracting the final mask from GPU, host-side "pinned" memory needs to be used instead of the slower pageable memory [99].

## III. METHODS

In this study, we target low-NA (0.33) EUV, single exposure, patterning of 28-36 nm pitch M0 and M2 designs in IMEC's 3 nm node. Because high-NA EUV is not yet ready for high-volume manufacturing, single-patterning of tight-pitch lower metal layers is essential to avoid the cost and complexity of multi-patterning [101,102]. We used low-n, bright field mask lithography with metal oxide resist, previously demonstrated at IMEC [101]. We corrected a total of 25 layouts covering random-logic and SRAM patterns, tip-to-tip spacings of 14-28 nm, and line CD biases of 0-2 nm. Overall, both CPU baseline and cuLitho OPC flows corrected 1.6 billion polygons in total, spanning over an area of 22 mm². Baseline runs had only on-focus correction, as through-focus OPC is prohibitively expensive on CPUs. In contrast, cuLitho run times were measured with and without through-focus correction. An additional set of experiments employed an AI-based OPC method, which further accelerates the cuLitho flow. Details of all methods are provided later in this section.

The OPC model, featuring freeform source, was calibrated for exposure on ASML NXE:3400B EUV scanner under both on-focus and through-focus conditions. No SRAFs were used because of the risk of their printing at the minimum width allowed by MRC. We used cuLitho-accelerated Proteus OPC engine from Synopsys to execute the correction flow [103]. We used NVIDIA Hopper (H200) GPU with HBM3e memory to run cuLitho accelerated flow and all the speedup data mentioned in this paper are on this machine [104]. The corrected masks were validated through EPE measurements on simulated wafer contours. Silicon validation data was collected after resist development (ADI) using Hitachi-GS1000 SEM metrology tool. A total of 7,350 SEM images were collected across the layers for defect inspection and CD measurements.

## A. cuLitho OPC flow

We use model based OPC for correction where the layout is segmented and corrections are applied by iteratively moving the segment positions. The iterative flow proceeds as follows:

1. Prediction: The imaging model runs to predict the contour based on the current mask geometry.
2. Error Calculation: On-focus and off-focus Edge Placement Errors (EPE) are calculated. These are then used for through-focus correction to yield the effective EPE and the Mask Error Enhancement Factor (MEEF).
3. Propose segment moves: The effective EPE and MEEF are used to propose new segment moves for the mask.
4. Rule Check: Mask Rule Checks (MRC) are performed on the proposed new mask geometry.
5. Acceptance and iteration: Only segment moves that satisfy the MRC checks are accepted, resulting in the final geometry for the current iteration's mask. This new mask geometry is then used to generate the aerial image for the next iteration of the solver.

The calculation of aerial image intensity $I(r; F, D)$ at a specific wafer position ($r$) under given process conditions focus ($F$) and dose ($D$) for partially coherent illumination is governed by Hopkins equation (Eqn. (1)). To accurately incorporate the thick mask effects, we employ a hybrid Electro Magnetic (EM) and imaging pipeline. This approach preserves the physics of thick-mask while also making it computationally tractable. Specifically, the mask is treated as a near-field electromagnetic problem and the resulting frequency domain mask response $O_{EM}(f, \sigma, p)$, which includes the 3D mask effects, is computed via EM methods [17,18]. Crucially, the Transmission Cross Coefficient (TCC) is separated from the thick mask effects to significantly reduce the computational cost [105,106].

$I(r; F, D) =$
$D \iint \sum_p O_{EM}(f_1, \sigma, p) O_{EM}^*(f_2, \sigma, p) TCC(f_1; f_2; F) e^{(2\pi r(f_1 - f_2))} df_1 df_2$      (1)

where:
F is focus and D is dose process condition
$f_1$ and $f_2$ are spatial frequencies on the mask
$\sigma$ is the source coordinate
$p$ is polarization
$O(f)$ is complex object spectrum
$TCC(f_1, f_2)$ is transmission cross coefficient (TCC) in frequency domain which includes source, pupil, aberrations, coherence.
$O^*(r)$ is a complex conjugate.
In the Hopkins frequency-domain formulation, TCC written explicitly in terms of the source and the pupil as:

$$TCC(f_1, f_2; F) = \int S(\sigma) P(f_1 + \sigma; F) P^*(f_2 + \sigma; F) d\sigma \quad (2)$$

where:
$S(\sigma)$ is the normalized source intensity distribution (partial coherence function)

$P(f)$ is the pupil function of the projection lens (includes NA cut-off and aberrations).
TCC is Hermitian and positive semi-definite because of which it can be simplified with an Eigenvalue (SVD) decomposition as:

$$TCC(f_1, f_2) = \sum_{k=1}^{K} \lambda_k \Phi_k(f_1) \Phi_k^*(f_2) \quad (3)$$

where:
$\Phi_k$ represents coherent imaging modes
$\lambda_k$ are mode weights

Combining the decomposed TCC with Hopkins equation gives the thick-mask aerial image as a sum of coherent systems (Eqn. (4)). A coherent image calculation is the convolution of near-field electromagnetic effects including full 3D mask effects ($O_{EM}$) and coherent imaging modes ($\Phi_k$).

$$I(r; F, D) = \sum_k \lambda_k |O_{EM} \otimes \Phi_k|^2 \quad (4)$$

The wafer contour is captured using an effective resist model in which the Hopkins aerial image is convolved with a calibrated resist kernel ($K_{resist}(r)$) and thresholded ($T_{eff}$) to extract the resist edge. This model correlates well with silicon as it captures acid diffusion, PEB blur and resist bias.

$$C = \{r \mid I(r; F, D) \otimes K_{resist}(r) = T_{eff}\} \quad (5)$$

The Edge Placement Error ($EPE(r)$) measures how far the effective latent image is from the target image $E_{tgt}(r)$. With each iteration the OPC solver's objective is to minimize $EPE(r)$. In our flow, this computation for each segment of the mask is performed in parallel on GPU.

$$EPE(r; F, D) = -\frac{(E_{eff}(r; F, D) - E_{tgt}(r))}{\|\nabla E_{eff}(r)\|} \quad (6)$$

In cuLitho flow, we perform through-focus correction by calculating $EPE(r; F, D)$ for 3 focus conditions: 1) on-focus ($f_0$) 2) positive de-focus ($f_p$), and 3) negative de-focus ($f_n$). We get the effective Edge Placement Error ($EPE_{eff}$) to balance out on-focus and off-focus errors (Eqns (7)-(8)). Through-focus correction enlarges the process window by reducing the sensitivity of image fidelity to process variations, thereby improving manufacturing yield. We use a flexible-window approach where we sacrifice on-focus to help off-focus. However, if the on-focus is not within the tolerances the correction for on-focus error takes precedence over average error.

$$EPE_{avg} = average(EPE(r; f_0, D), EPE(r, f_p, D), EPE(r, f_n, D)) \quad (7)$$

$$EPE_{eff} = \begin{cases} EPE(r; f_0, D) - tol_{pos} \text{ if } EPE(r; f_0, D) > tol_{pos} \\ EPE(r; f_0, D) - tol_{neg} \text{ if } EPE(r; f_0, D) < tol_{neg} \\ EPE_{avg} \text{ if } tol_{neg} < EPE(r; f_0, D) < tol_{pos} \end{cases} \quad (8)$$

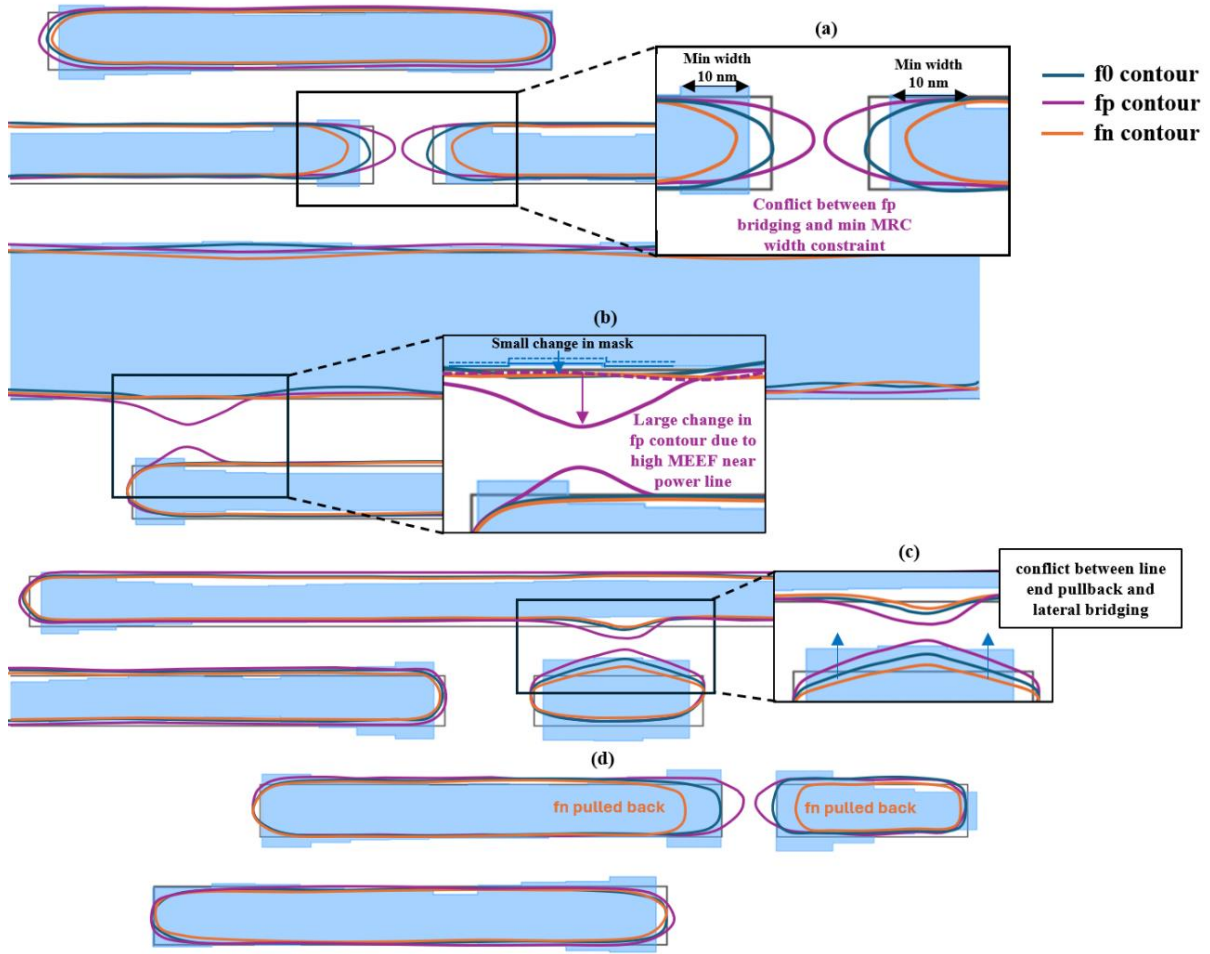

**FIG. 3.** Examples showing the challenges involved in through-focus OPC correction. Several context-aware strategies were used to address the following conflicts. a) Conflict between fp bridging and MRC min width constraint b) High MEEF near the power-line causing all the contours to bridge under small mask movements c) Conflict between line-end pull back and lateral bridging in short metals d) fn contour pulled back at isolated tip-to-tip regions.

The Mask Error Enhancement Factor (MEEF) is a crucial computation in OPC. MEEF quantifies the sensitivity of the contour to changes in the mask. Specifically, it is defined as the ratio of the change in the contour to the change in the mask shape resulting from mask segment movements. We use a controller scheme as a corrector where the calculated MEEF is then used as a feedback. This dynamic feedback controller approach significantly reduces the number of iterations required for OPC convergence [107,108].

cuLitho addresses the various challenges in lithography, illustrated in FIG. 3, by applying different correction strategies during through-focus correction (TFC). These strategies include context-aware tolerances, design retargeting, and lateral segment movement restriction, which help resolve conflicts between different focus conditions and MRC constraints. Section IV B provides a detailed explanation of these strategies.

## B. AI in cuLitho OPC

Machine learning (ML) and AI have been increasingly applied to OPC. Early models employed regression [109,110], multilayer perceptrons [111], support vector machines [112], and Bayesian inference [113] to approximate mask corrections. More recent approaches leverage deep learning and generative AI to model the nonlinear design-to-mask transformation with high fidelity, offering speedups by serving as learned surrogates for computationally expensive physics simulations [114,115].

Generative models, particularly conditional generative adversarial networks (GANs), have demonstrated the ability to learn pixel-aligned mappings between design and mask spaces. However, existing GAN-based mask-optimization studies primarily attempt to directly mimic final OPC or ILT masks without incorporating lithography simulations to guide the AI model [116]. Furthermore, their objective functions typically include only image-matching L1/L2 losses, lacking differentiable lithography-specific terms capable of updating model weights through backpropagated gradients. As a result, prior methods may perform acceptably on in-sample patterns or small clips with a few polygons for academic purposes but fail to generalize for full-chip industry applications due to limited extrapolation capability. Many also produce curvilinear masks incompatible with Manhattan-based mask rule checking (MRC) and require additional post-processing to enforce MRC cleanliness, negating much of the runtime benefit gained from AI-generated masks.

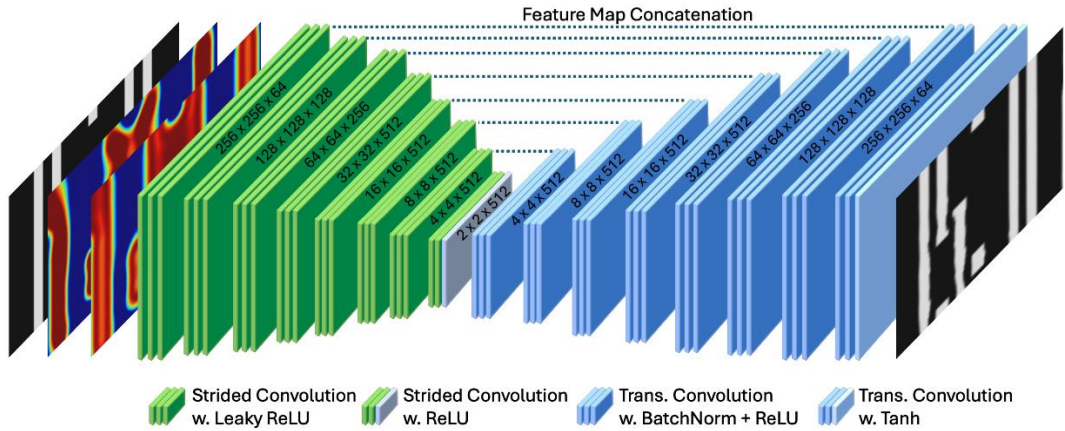

**FIG. 4.** Architecture of the proposed deep encoder–decoder network.

Our method accelerates OPC without modifying mask representations, mask rules, data structures, or OPC engines. We achieve this via a lithography-guided conditional GAN-UNet that outputs segment movements (Δs) rather than pixel-level masks. Combined with cuDOP-generated aerial/gradient context signals and full-stack cuLitho acceleration, this approach produces manufacturable, infrastructure-compatible AI initialization that significantly reduces OPC convergence time while ensuring lithographic plausibility.

We construct a three-channel input tensor by concatenating physically meaningful quantities, each scaled by a global normalization factor so that all values lie within [0,1]:

- $M^{(0)} = Z_t$: target design
- $I^{(0)}$: aerial intensity of the target design
- $\frac{\partial I}{\partial M}$: intensity gradient with respect to the mask

The resulting input tensor $X$ is

$$X = concat\left(M^{(0)}, I^{(0)}, \left(\frac{\partial I}{\partial M}\right)^{(0)}\right) \in [0,1]^{H \times W \times 3} \quad (9)$$

A conventional OPC flow is used to generate the corrected mask. After 16 OPC iterations followed by MRC checks, the final polygonal mask $M^{(16)}$ is rasterized using cuDOP's mask rasterizer and converted into a single-channel image normalized to [0,1] for the output $Y$.

$$Y = M^{(16)} \in [0,1]^{H \times W \times 3} \quad (10)$$

The generator $G_\theta$ is implemented as an 8-level UNet designed to capture both fine-grained mask details and long-range lithographic context. The architecture consists of:

- 64 base convolutional filters, doubled at each encoder depth, capped at 512 filters
- LeakyReLU activations in the encoder

- ReLU activations in the decoder
- Skip connections at all spatial scales to preserve high-frequency mask features
- Final output: a continuous-valued mask field

In the architecture of deep encoder-decoder network used in the current work as shown in FIG. 4, strided convolutions with Leaky ReLU are used in the encoder, and transposed convolutions with batch normalization and ReLU are used in the decoder. Skip connections link corresponding resolution levels to retain spatial detail and improve reconstruction fidelity.

We get a differentiable approximation of the OPC-updated mask from the generator as given in Eqn. (11). This structure provides a balance between expressive capacity and stability during adversarial training.

$$M_c = G_\theta(X) \quad (11)$$

The discriminator $D_\phi$ adopts a standard PatchGAN formulation, operating on local mask patches. It is trained jointly with the generator to enforce local realism and lithographic consistency in the predicted mask field. By evaluating patches rather than the full image, PatchGAN ensures that the generator produces spatially coherent mask patterns aligned with manufacturable OPC corrections.

The generator is trained using a composite loss that combines direct mask supervision, adversarial realism, and lithography-consistent printing behavior. The overall objective is given below:

$$L = \lambda_1 L_{mask} + \lambda_2 L_{GAN} + \lambda_3 L_{litho} \quad (12)$$

where:
$L$ is the total loss function
$L_{mask}$ is the mask loss function
$L_{litho}$ is lithography-consistency loss

In the current work, the three loss functions are calculated in Eqns. (13)-(15) using continuous generator output ($M_c$),

and converged OPC mask obtained after 16 full iterations including MRC corrections ($M^{(16)}$).

The primary supervision signal is a direct match between the generated mask and the converged OPC mask:

$$L_{mask} = \left|\left| M_c - M^{(16)} \right|\right|_1 \tag{13}$$

This forces the generator to approximate the final manufacturable mask produced by the conventional OPC flow and ensures that GAN training is grounded by physically meaningful supervision.

A PatchGAN discriminator $D_\phi$ is trained to distinguish converged OPC masks from AI-generated masks. Using a least-squares formulation for stability the adversarial loss is given in Eqn. (14). $E_{M^{(16)}}$ is average over real data samples drawn from real data distribution and $E_X$ is average over generated samples drawn from model distribution. This encourages the generator to produce mask features that are locally indistinguishable from those generated by a full OPC engine.

$$L_{GAN} = E_{M^{(16)}}\left[\left(D_\phi\left(M^{(16)}\right) - 1\right)^2\right] + E_X\left[D_\phi(M_c)^2\right] \tag{14}$$

To ensure that the AI-generated mask actually prints the intended design, we compare two binary printable images calculated in Eqns. (16)-(17). The lithography-consistency loss is defined in Eqn. (15), this term penalizes the deviations between what the AI-generated mask prints ($Z_{print}$) and what the lithography system can print from the target design ($Z_{round}$).

$$L_{litho} = \left|\left| Z_{print}\left(M_c\right) - Z_{round} \right|\right|_1 \tag{15}$$

The ideal layout $Z_t$ is passed through a Gaussian low-pass filter ($G_\sigma$), reflecting the resolution limits ($\tau_{round}$) of the optical system. This produces a smoothed ("rounded") target that reflects the contours the scanner can reproduce.

$$Z_{round}(x, y) = 1 \; for \; \{(G_\sigma Z_t)(x, y) \geq \tau_{round}\} \tag{16}$$

The print from the AI-generated mask is computed using a lithography-relevant threshold ($\tau_{print}$) as shown below:

$$Z_{print}(x, y; \; M_c) = 1 \; for \; \{I(M_c)(x, y) \geq \tau_{print}\} \tag{17}$$

Given the depth of the proposed neural network, controlling overfitting is essential. Accordingly, both in-sample (training) and out-of-sample (validation) losses are monitored throughout the training process. As shown in FIG. 5, the in-sample loss exhibits an overall decreasing trend with minor oscillations, while the out-of-sample loss decreases initially, reaches a minimum, and subsequently increases, indicating the onset of overfitting. The final model is therefore selected at the epoch corresponding to the minimum out-of-sample loss, effectively employing an early-stopping criterion to enhance generalization performance.

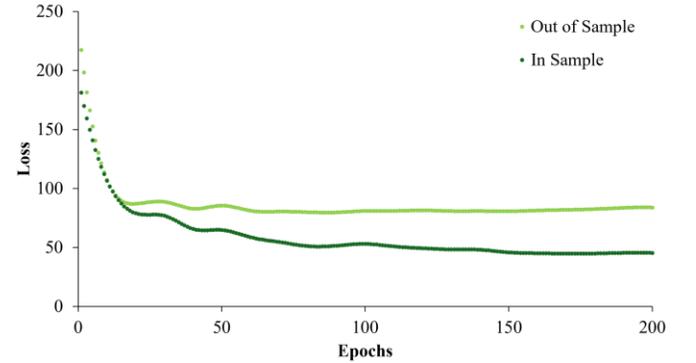

**FIG. 5.** Out of sample and in sample losses measured throughout the training process

Unlike approaches that produce curvilinear or purely rasterized masks, our framework maintains full compatibility with Manhattan-style OPC flows by converting the continuous generator output into segment-level movements applied directly to the original polygonal layout. This ensures manufacturability, preserves layout topology, and allows seamless adoption within existing OPC infrastructure.

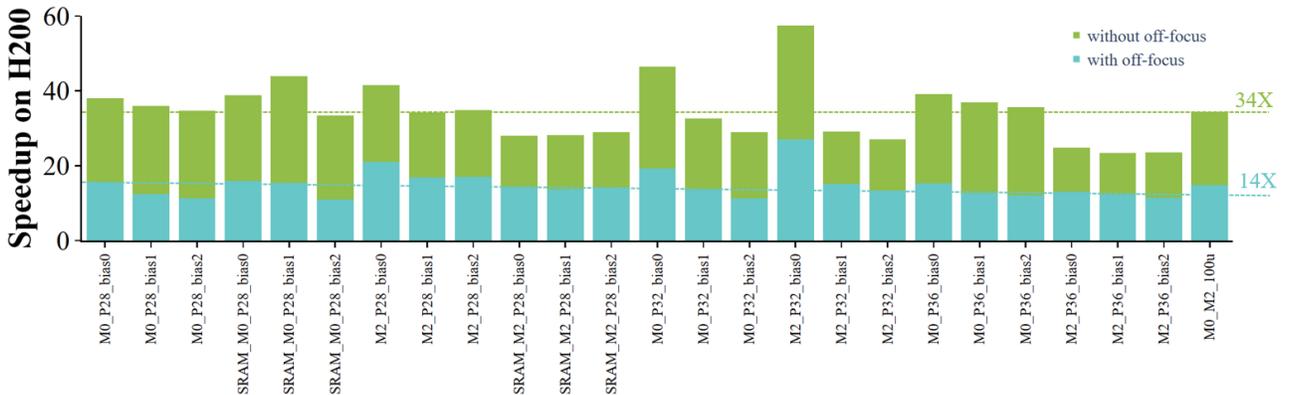

**FIG. 6.** cuLitho speedups – ratio of CPU baseline run time to cuLitho run time for various layouts. On average, cuLitho is 34X faster than CPU without through-focus correction. Speedup remains substantially high at 14X after introducing through-focus correction.

The workflow used in the current work is as follows:

- Binarization of the generator output: Threshold the continuous mask prediction $M_c$ to obtain a clean binary mask suitable for contour extraction.
- Contour extraction: Apply marching squares to the binarized mask to polygonize the predicted mask boundary.
- Segment-wise displacement measurement: For each Manhattan segment of the original design $M^{(0)}$, find the corresponding portion of the extracted contour and measure the normal displacement using cuLitho's measurement APIs. The resulting scalar movement is denoted $\Delta s_i$.
- Construction of AI-initialized mask: The initial mask used for downstream OPC iterations is obtained by applying these movements to the original segments ($M_{init} = M^{(0)} + \Delta s$).

With the above mentioned approach, we guarantee the following:

- Manhattan topology preserved, no curvilinear edges are introduced.
- MRC constraints remain valid, no new jogs, slivers, acute angles, or non-Manhattan geometries appear.
- Full compatibility with commercial OPC flows, the mask maintains its standard format, polygon structure, and constraint set.
- No modifications required to rule decks, file formats, or downstream signoff infrastructure.

This segment-movement representation is a practical innovation of our method, enabling the AI-generated initialization to plug directly into production OPC pipelines without altering existing manufacturing rules. The AI-initialized mask integrates directly with the end-to-end GPU-accelerated computational lithography stack. By providing a high-quality initial condition for the mask from AI, the OPC iteration count is reduced by 2X, providing a significant speedup boost.

## IV. RESULTS AND DISCUSSION

### A. Speedup from AC and AI

FIG. 6 shows the measured cuLitho speedup, defined as the ratio of CPU baseline runtime to cuLitho runtime for different layouts. Without through-focus correction, cuLitho using accelerated computing (AC) delivers an average 34X speedup relative to the CPU baseline. FIG. 6 also illustrates the impact of incorporating through-focus correction. As anticipated, the additional computational burden of through-focus modeling reduces the speedup relative to the on-focus-only CPU baseline. Despite this overhead, cuLitho with through focus OPC correction still maintains a substantial 14X speedup compared to the CPU baseline.

As mentioned earlier, the AI model provides superior initial conditions for the OPC correction loop, reducing the number of iterations required for convergence by half. Consequently, AI delivers an additional 1.7X performance gain over cuLitho-AC, as shown in FIG. 7. For the case without through-focus correction, the average speedup increases from 34X to 57X due to the boost from AI. A similar performance enhancement was observed for flows that include through-focus correction, increasing their average speedup from 14X to 24X. We also observe that the correction from AI does not sacrifice fidelity (EPE) or MRC constraints as shown in FIG. 8. Note that AI model training time is excluded from this comparison, as training represents a one-time upfront cost. Once trained, the model can be reused for all subsequent tapeouts employing the same OPC model and layout type.

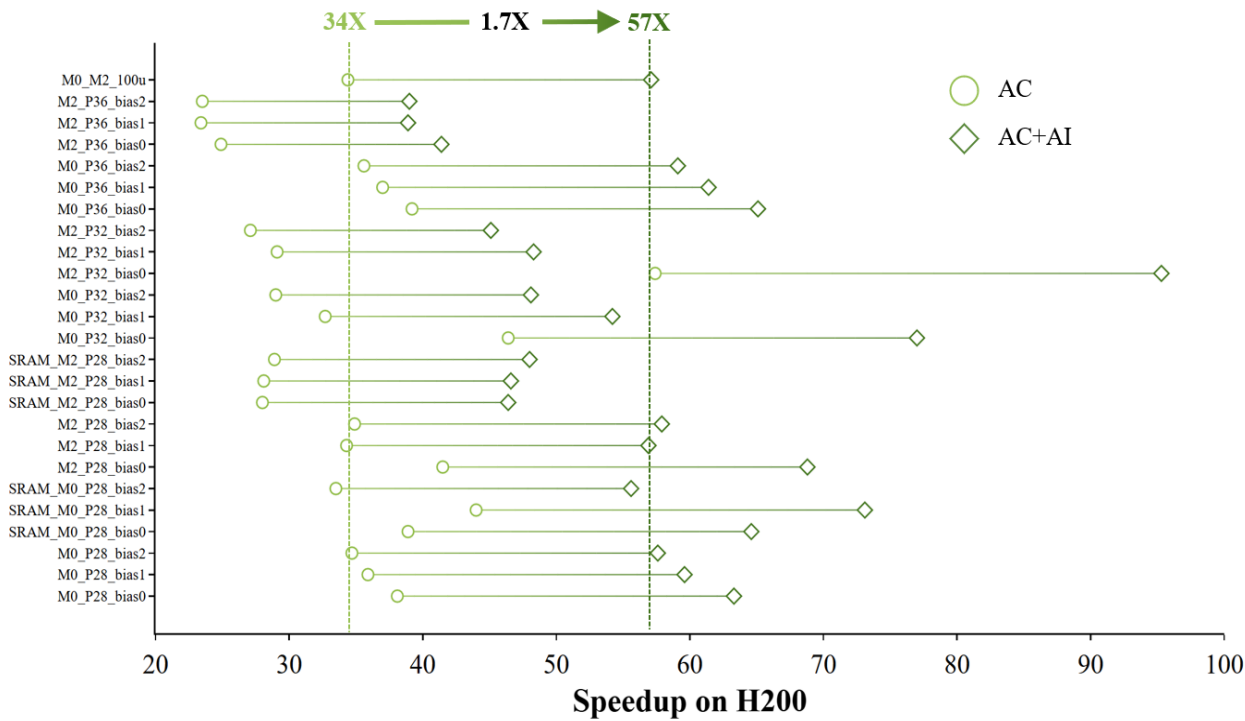

**FIG. 7.** Performance boost from AI for various layouts. On average, AI provides an additional 1.7X speedup over AC.

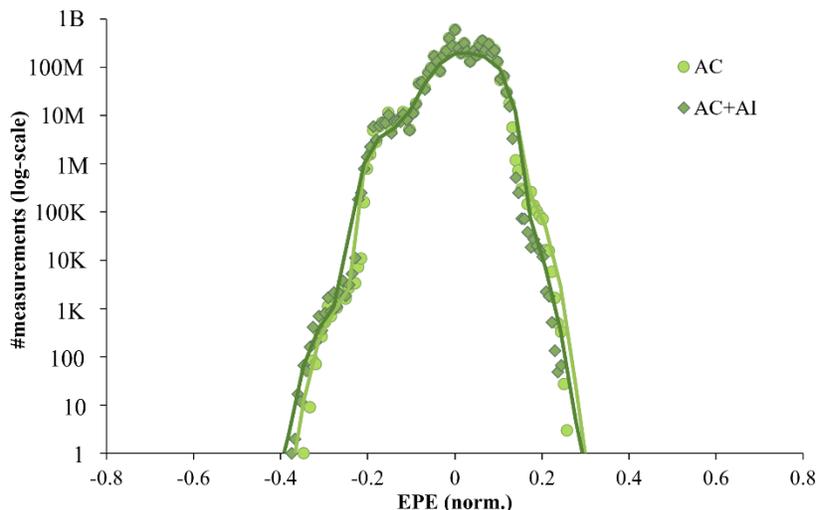

**FIG. 8.** AI provides additional 1.7X speedup without sacrificing fidelity and matching EPE with AC. The masks corrected by AI (192 million polygons total) satisfied all 7 MRC constraints.

## B. Pre-Silicon validation

We validated the final cuLitho mask using Mask Rule Check (MRC) and simulated wafer Edge Placement Error (EPE) checks to ascertain its manufacturability and lithographic quality. FIG. 9 shows the pre-silicon MRC and fidelity verification results. The MRC cleanliness of all the masks was verified across eight rules – spacing, width, internal and external corner, notch, nub, jog distances and polygon area – and across 200 billion edge interactions. The lithographic quality of the mask was determined using simulated EPE checks at best focus, positive defocus, and negative defocus conditions. As seen in FIG. 9, the through-focus correction cuLitho OPC resulted in an overall EPE improvement of 1.6X. In particular, the peak best-focus EPE improved from 4.4 nm for CPU to 2.1 nm for cuLitho, and the peak defocus EPE improved from 7 nm to 4.1 nm. The peak EPE is from challenging tight-pitched configurations for single print EUV lithography [101].

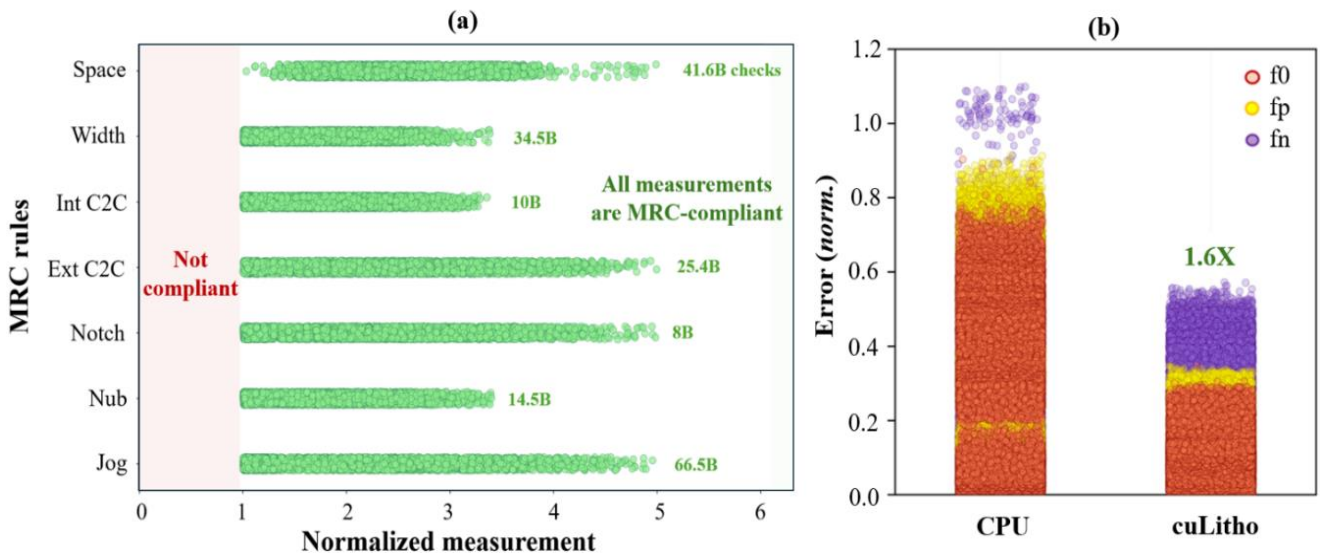

**FIG. 9.** Pre-silicon validation of cuLitho corrected masks. (a) MRC cleanliness: measured distances (normalized by MRC tolerance) and total check counts across all cuLitho corrected masks for each MRC rule. All masks are MRC compliant. (b) Normalized through-focus EPE measured from simulated wafer checks. cuLitho correction improves EPE error by 1.6X, on average

The following set of specialized, context-aware OPC strategies were used to ensure the lithographic fidelity of cuLitho masks.

1. *Enhanced Line-End Correction using Hammer-Heads:* To improve line-end fidelity, an initial mask shape incorporating hammer-head shaped metal lines was employed. This initial strategy proved effective in reducing the Edge Placement Error (EPE) at line ends under the negative-focus (fn) condition without leading to bridging in the positive-focus (fp) condition for tighter configurations.

2. *Global Retargeting:* To proactively prevent the high-risk scenario of bridging between opposing line ends (tip-to-tip locations), the original design was systematically retargeted. This involved pulling the line ends inward by a minimal yet effective distance of 1 nm. This slight inward movement provides an immediate, robust safety margin against the contour expansion that causes bridging under positive defocus.

3. *Context-Aware MEEF Driver Control:* Isolated tip-to-tip locations are particularly susceptible to the detrimental effects of TFC, showing extreme sensitivity to focus variations. To specifically address this sensitivity and balance the corrections, a sophisticated context-aware Mask Error Enhancement Factor (MEEF)-based driver was employed. The MEEF-based driver relates mask correction to the resulting wafer change. Crucially, the negative-focus (fn) component of this driver was removed. By taking out the fn driver, the system shifts its corrective focus. This prevents over-correction aimed at negative focus (which could lead to pinching) and thus successfully prevents the primary failure mode which is positive-focus bridging that often results from the attempt to aggressively correct for the negative focus condition.

4. *Power Line Exception (Sensitivity Exclusion):* Not all features can be subjected to the same correction drivers and retargeting schemes. A specific exception was created for metal lines located in the vicinity of long, 20 nm wide power lines within the M0 layer. For these lines, both MEEF-based driver corrections and retargeting were intentionally omitted. This exclusion is critical because the electromagnetic and physical environment of this specific region makes it acutely sensitive. Minor, necessary mask alterations introduced by OPC tend to produce undesirable and disproportionately large effects on the wafer in this area, making a "hands-off" approach the most stable solution.

5. *Lateral Segment Movement Restriction:* Lateral bridging, or bridging between the side segments of neighboring lines, poses another significant risk, particularly for short metal lines (defined here as those less than 50 nm in length). To eliminate this issue, the adjustment of lateral segments was strictly limited. This restriction specifically prohibits the use of lateral segment movement as a means to correct for End-of-Line Edge Placement Error. Allowing lateral segment movement for line end correction inherently causes those segments to move outward (laterally), directly leading to the formation of the critical lateral bridging failure with adjacent lines. By imposing this limit, the short lines maintain their design separation integrity.

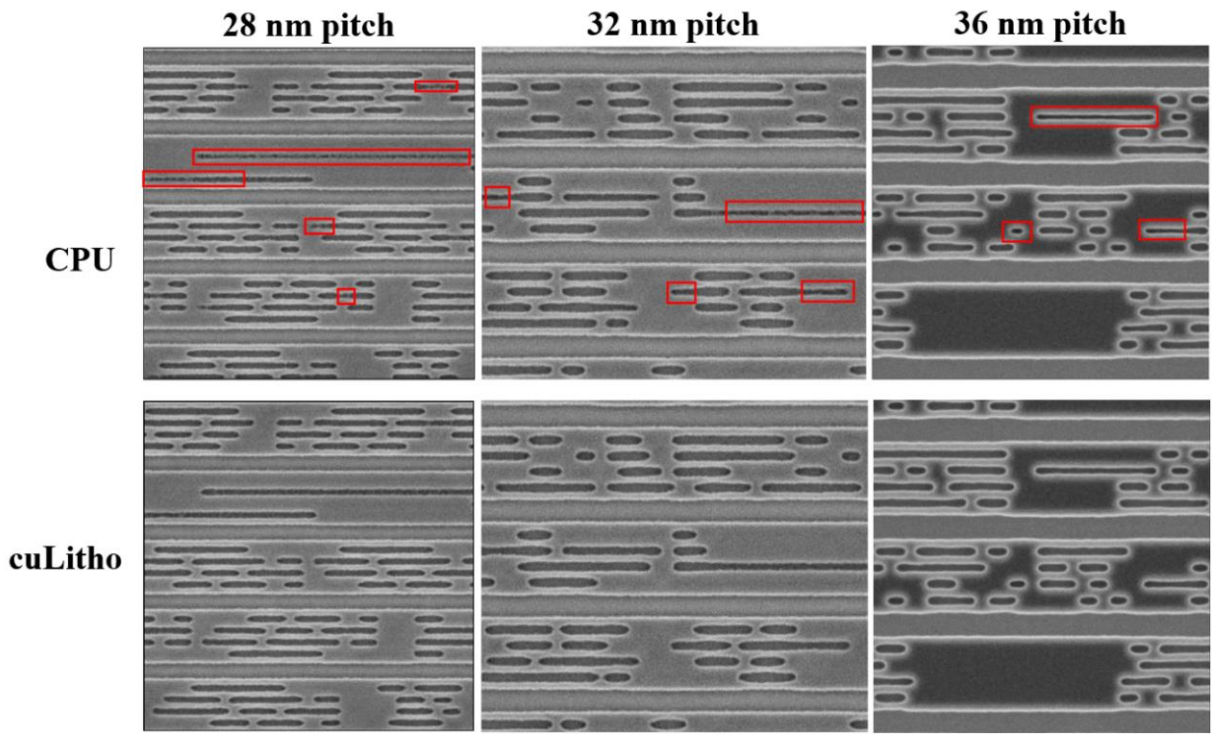

**FIG. 10.** Comparison of SEM wafer images corresponding to CPU and cuLitho masks for M0 P28, P32 and P36 layouts at the -45 nm defocus condition. Across the pitches, through-focus cuLitho OPC prevents isolated and semi-isolated line pinches seen in CPU patterning (red rectangles).

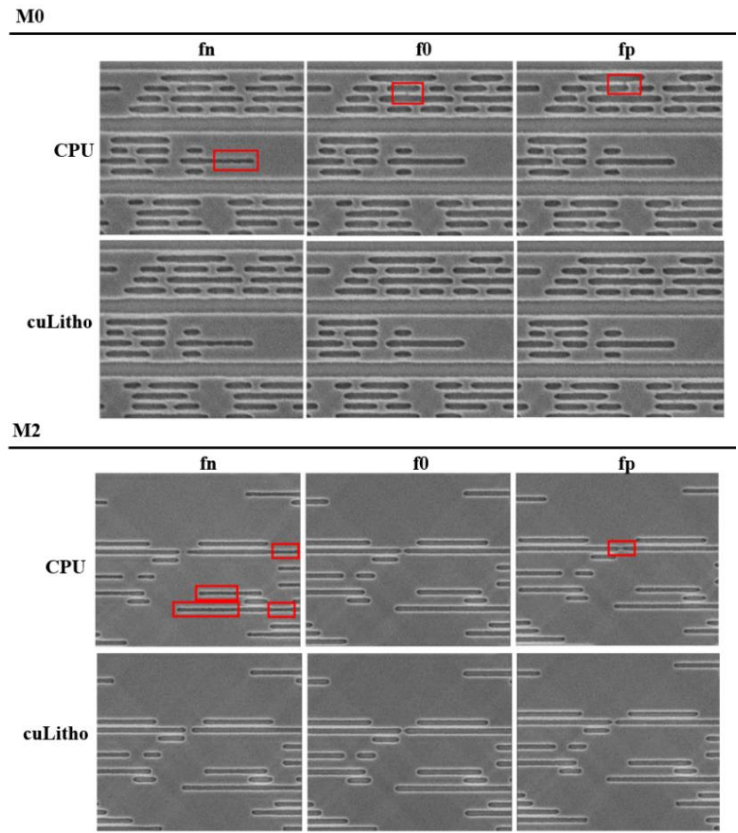

**FIG. 11.** Comparison of through-focus SEM wafer images corresponding to CPU and cuLitho masks for M0 and M2 designs. CPU patterning fails at both negative (fn) and positive defocus (fp) conditions for both M0 and M2 designs, as indicated by red rectangles. Failure at negative defocus is due to pinching of isolated and semi-isolated lines. In contrast, failures at positive defocus are due to lateral or tip-to-tip bridging. cuLitho improves through-focus patterning while maintaining the fidelity at best focus (f0).

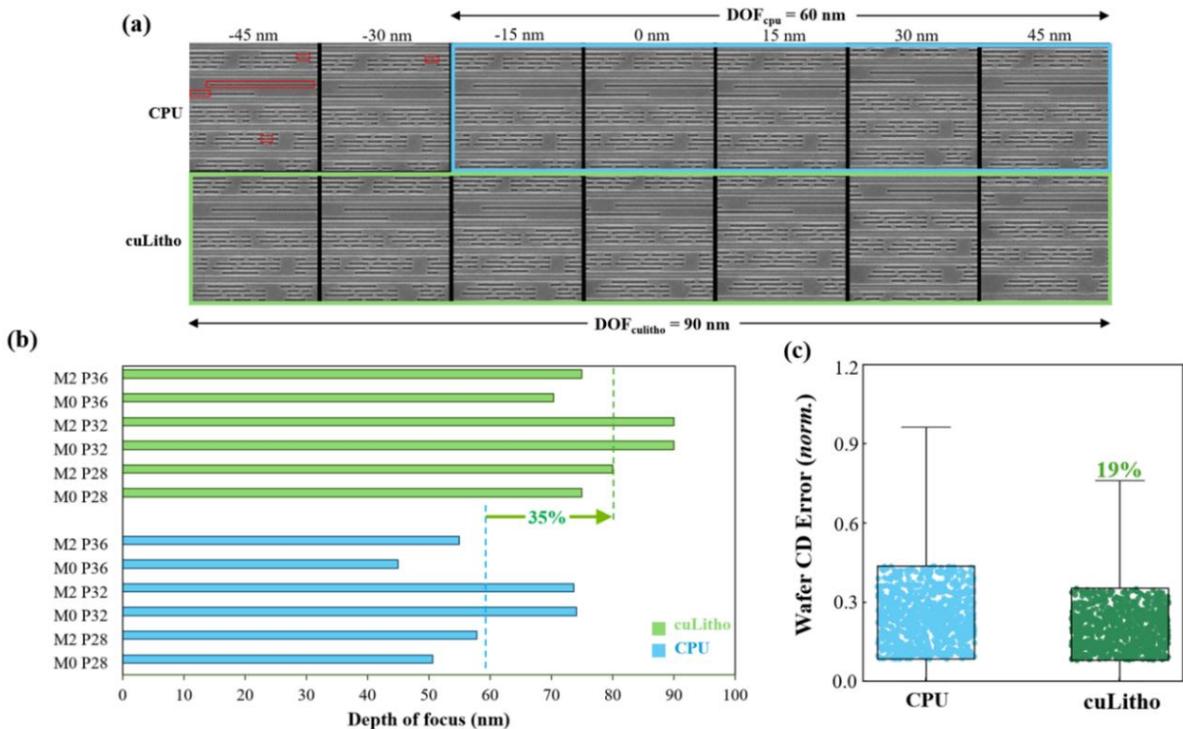

**FIG. 12.** On-silicon depth-of-focus (process window) and wafer CD error comparisons between CPU and cuLitho masks. (a) illustration of depth-of-focus extraction from SEM images, (b) measured depth-of-focus across all layouts, and (c) measured wafer CD error across all layouts. Through-focus cuLitho correction improves depth-of-focus by 35% and wafer CD error by 19% compared to the CPU baseline.

## C. Silicon validation

We focus on random-logic layouts for silicon validation, our main interest is the through-focus behavior at best dose, the conditions for which we have optimized the cuLitho mask. For defect inspection and Critical Dimension (CD) measurements, a total of 7,350 SEM images were collected. This sample included different M0 and M2 layers across 28 nm, 32 nm, and 36 nm pitches. We used a comprehensive sampling strategy, focusing on the most representative and challenging configurations like tip-to-tip, power line neighborhoods, and isolated locations. To minimize noise, the raw SEM images were averaged over five repeated captures for the images presented here. Under the best focus conditions, both CPU and cuLitho masks achieved comparable patterning quality. However, as illustrated in FIG. 10 for the M0 design, CPU patterning fails at -45 nm defocus due to shrinking or pinching of isolated and semi-isolated lines. In contrast, the cuLitho mask maintains fidelity across all three pitches due to its effective through-focus correction. At positive defocus, cuLitho correction prevents tip-to-tip and lateral bridging violations, as shown in FIG. 11. The figure demonstrates that these observations extend to M2 designs. The enhanced variation resilience of cuLitho correction is further evidenced by the depth-of-focus (process window) comparison shown in FIG. 12. On average, cuLitho correction expands depth-of-focus by 35% compared to the CPU baseline. We also extracted 1,463,229 wafer CD data measurements from the images and computed the CD error – defined as the absolute difference between the target design CD and the actual measured CD on silicon. We find that cuLitho correction reduces through-focus CD error by 19% on average across all layouts compared to the CPU-generated mask, as shown in FIG. 12.

## V. BROADER IMPACT

Beyond the faster cycles from the dramatic 57X acceleration and inclusion of more accurate lithography solutions (variation-aware OPC as an example demonstrated in this paper), this work has broader implications.

1.  This work represents a significant step towards improving the sustainability of semiconductor computing. We evaluate the cost efficiency and carbon footprint of cuLitho-accelerated OPC using NVIDIA H200 GPUs relative to a CPU implementation on AMD Milan processors. Our analysis assumes a production workload of 30 full-reticle layers processed per day in a fabrication environment, and is evaluated over a three-year period. We compare infrastructure scale, power consumption, cost per layer processed, and $CO_2$ emissions. With a 57X speedup, 353 H200 HGX systems deployed across 118 power-constrained racks deliver the target throughput, replacing 40,690 CPU servers across 1,403 racks. This corresponds to an approximately 12X reduction in both server count and rack footprint. Power consumption is reduced from 37.6 MW in the CPU-only datacenter to 2.9 MW in the GPU-based deployment, reflecting the substantially higher computational density of GPUs. These infrastructure gains translate directly into economic and environmental benefits. Considering compute costs alone, the GPU-based system achieves a 5.7X reduction in cost per layer processed relative to the CPU baseline. Normalized per layer, $CO_2$ emissions are reduced by 12.9X, primarily due to lower power demand and a significantly smaller datacenter footprint. Overall, GPU-accelerated OPC enables high-throughput computational lithography with substantially lower cost and carbon intensity than traditional CPU computing.

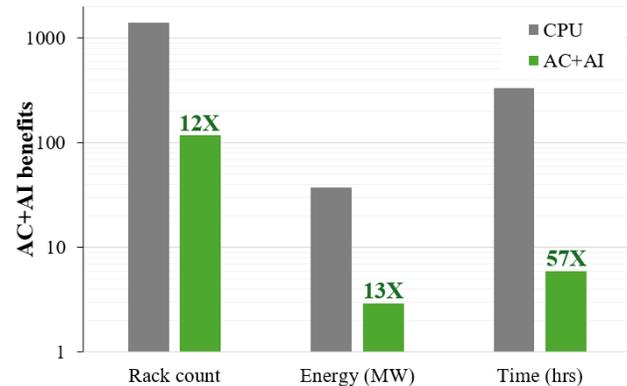

**FIG. 13.** Leveraging AC and AI in OPC achieves a 12X reduction in rack count, 13X energy savings, and a 57X decrease in compute time.

2.  The primitives accelerated in this work extend to other compute-intensive semiconductor workloads.
    a.  Physical design verification: Design Rule Checking (DRC) is an indispensable step ingrained in every stage of modern chip design, bridging design with manufacturing. With process nodes nearing the limits of physics, design rules have grown exponentially, increasing both design-cycle times and compute demands. DRC involves geometric operations on billions of polygons, including Boolean primitives, connectivity queries, proximity analysis, and pattern matching. cuLitho primitives are tailor made for DRC flows – promising dramatic acceleration of DRC runtimes, from days to hours. This will enable significantly faster iterations, shorter fab tapeout schedules, and faster times-to-market.
    b.  Mask and wafer inspection: Mask and wafer inspection pipelines increasingly rely on computationally intensive image-processing and modeling tasks. These include aerial image simulation, defect detection, hotspot classification, and forward/adjoint modeling for nuisance-reduction. cuLitho accelerates these workloads by providing high-throughput GPU primitives for computational geometry, optics, and stencil-like operations that dominate inspection pipelines. GPU-optimized kernels dramatically reduce latency for image formation, edge extraction, and frequency-domain filtering, while its batched, streaming execution model keeps device utilization high for

large-volume inspection scans. As inspection shifts toward higher numerical aperture systems and angstrom-scale patterning tolerances, the computational load grows exponentially; cuLitho scalable primitives offer the throughput required for real-time or near-real-time defect analysis, thereby enabling more rigorous detection thresholds without compromising throughput.

## VI. CONCLUSION

The twin technologies of accelerated computing and AI are reshaping scientific computing. This work extends them in the field of chipmaking by targeting its largest workload, computational lithography. This new computing platform necessitates a fundamental reinvention of the full computing stack, driven by the unforgiving nature of Amdahl's law. cuLitho redesigns the building blocks and accelerates them in the range from 50X to 1500X. In this work, we also developed a diffractive-optics-informed generative-AI model to predict near-perfect masks – cutting down iterations by half. Combined, they achieved 57X speedup on 25 EUV designs from IMEC – with no loss in accuracy. cuLitho masks met all fidelity targets: mean OPC error below 0.2nm and MRC clean outputs across 200 billion inter- and intra-polygon interactions. Furthermore, the 57X speedup translates directly to a 13X reduction in computational emissions, marking a significant step toward sustainable computing in the semiconductor industry.

The expanded compute and power envelope also enables inclusion of rigorous solutions and next-generation technologies – computationally prohibitive thus far. To that end, we reinvested a fraction of the freed-up compute to incorporate variation-aware OPC in our simulations. Mask fidelity improved by 1.6X: peak EPE (occurring at the tightest most challenging tip-to-tip pattern configurations) improved by 2.1X for best-focus (4.4 nm vs 2.1 nm), and 1.7X for defocus conditions (7 nm vs 4.1 nm). While these simulation results were strongly indicative of cuLitho goodness, definitive validation came from silicon experiments at IMEC. On 1.5 million silicon measurements, cuLitho-based OPC showed significant benefits compared to conventional methods: 35% better process window (depth-of-focus improved from 59nm to 80nm) and 19% better through-focus wafer CD error (decreased from 3.7 nm to 3.0 nm) – all while staying 24X faster.

This work represents the most significant advance in computational lithography in over a decade, with 1) dramatically shortening design-to-production cycles, 2) enabling adoption of next-generation innovations, and 3) transitioning the semiconductor industry toward a sustainable computing paradigm. With its foundational primitives extending to other critical workloads – including DRC, and mask and wafer inspection – this is a first step towards a complete transformation of semiconductor computing.